\documentclass[12pt,twoside,english]{article}
\usepackage[T1]{fontenc}
\usepackage[latin1]{inputenc}
\usepackage{amstext}
\usepackage{amssymb}
\usepackage{babel}
\begin{document}

\title{Semiclassical treatment for molecular rotation spectra in high electric
fields}

\author{{\normalsize{M. Apostol$^{*}$ and L. C. Cune }}\\
{\normalsize{Department of Theoretical Physics, Institute of Atomic
Physics, }}\\
{\normalsize{Magurele-Bucharest MG-6, POBox MG-35, Romania }}\\
{\normalsize{$^{*}$corresponding author, email: apoma@theory.nipne.ro}}}

\date{{}}

\maketitle
\relax
\begin{abstract}
Molecular rotation spectra, generated by the coupling of the molecular
electric-dipole moments to an external time-dependent electric field,
are discussed in a few particular conditions which can be of some
experimental interest. First, the spherical-pendulum molecular model
is reviewed, with the aim of introducing an approximate method which
consists in the separation of the azimuthal and zenithal motions.
Second, rotation spectra are considered in the presence of a static
electric field. Two particular cases are analyzed, corresponding to
strong and weak fields. In both cases the dipoles perform rotations
and vibrations about equilibrium positions, which may exhibit parametric
resonances. For strong fields a large macroscopic electric polarization
may appear. This situation may be relevant for polar matter (like
pyroelectrics, ferroelectrics), or for heavy impurities embedded in
a polar solid. The dipolar interaction is analyzed in polar condensed
matter, where it is shown that new polarization modes appear for a
spontaneous macroscopic electric polarization (these modes are tentatively
called \textquotedbl{}dipolons\textquotedbl{}); one of the polarization
modes is related to parametric resonances. The extension of these
considerations to magnetic dipoles is briefly discussed. The treatment
is extended to strong electric fields which oscillate with a high
frequency, as those provided by high-power lasers. It is shown that
the effect of such fields on molecular dynamics is governed by a much
weaker, effective, renormalized, static electric field.
\end{abstract}
\relax

\emph{Key words:} rotation molecular spectra; high electric fields;
parametric resonance; spontaneous polarization; highly-oscillating
electric fields 

PACS: 33.20.Sn; 45.20.dc; 33.20.Xx; 42.65.Yj;77.80.-e 

\newpage

\noindent \textbf{Introduction.} Usually, the molecular dynamics
in the presence of static electric fields is limited to weak fields,
as those produced currently in the laboratory. A small, orientational
polarization of the electric-dipole moments is well known in this
case, governed by the Curie-Langevin-Debye law. Comparatively, more
information is available for the dynamics of the magnetic moments
in the presence of static magnetic fields available in the laboratory,
although the magnetic moments are much smaller than the electric-dipole
moments. Orientation, deflection, trapping of polar molecular beams
in static electric fiels are also known.\cite{key-1}-\cite{key-7}
Classical dynamics of polar molecules (rigid spatial rotators) in
combined static and non-resonant electric fields has also been studied
in Refs. \cite{key-8}-\cite{key-14} 

With the advent of high-power lasers, the interest for the molecular
dynamics in high electric fields may be revived. Although the electric
fields produced in the laser beams are oscillating in time, we show
in this paper that their effect on the molecular dynamics is that
of weaker, renormalized, static fields, as a consequence of their
much higher frequency in comparison with the molecular rotation or
vibration frequencies.

First, we review briefly the spherical-pendulum molecular model with
the aim of defining our working method, which consists in the separation
of the azimuthal and zenithal motions. The method is valid for heavy
molecules. Second, we apply this method to high electric fields, where
parametric resonances are highlighted in the molecular rotation spectra.
Similar results are briefly discussed for weak electric fields. Further,
the dipolar interaction is analyzed in polar matter, where it may
produce a spontaneous polarization. A continuous model is introduced
for the motion of this polarization, whose excitations are tentatively
called \textquotedbl{}dipolons\textquotedbl{}; it is shown that their
interaction with a time-dependent electric field may also exhibit
parametric resonances. Such arguments are briefly extended to similar
features exhibited by magnetic moments.

\textbf{Free rotations.} In many cases the free molecular rotations
are described satisfactorily by a spherical-pendulum model (spatial,
rigid rotator, spherical top).\cite{key-15,key-16} A spherical pendulum
consists of a point of mass $M$ which rotates freely in space at
the end of a radius \\
$\mathbf{r}=r(\sin\theta\cos\varphi,\sin\theta\sin\varphi,\cos\theta)$,
as described by the hamiltonian 
\begin{equation}
H=\frac{1}{2}M\dot{\mathbf{r}}^{2}=\frac{1}{2}Mr^{2}(\dot{\theta}^{2}+\dot{\varphi}^{2}\sin^{2}\theta)\,\,;\label{1}
\end{equation}
if the point has a charge $q$, then there is a dipole $\mathbf{d}=q\mathbf{r}$
which can couple to an external electric field $\mathbf{E}\cos\omega t$,
with an interaction hamiltonian $H_{int}(t)=-dE\cos\theta\cos\omega t$.
We take the electric field directed along the $z$-axis. As it is
well known, the hamiltonian given by equation (\ref{1}) can be writen
as $H=\mathbf{L}^{2}/2I$, where $\mathbf{L}$ is the angular momentum
and $I=Mr^{2}$ is the moment of inertia. The eigenfunctions are the
spherical harmonics, with the energy levels $E_{l}=\hbar^{2}l(l+1)/2I,\, l=0,1,2,...$. 

As it is well known, in the first-order of the perturbation theory
for the interaction $H_{int}(t)=-dE\cos\theta\cos\omega t$ the rate
of quantum transitions with frequency $\omega_{0}=(E_{l+1}-E_{l})/\hbar=(\hbar/I)(l+1)$
is 
\begin{equation}
\frac{\partial\left|c_{lm}\right|^{2}}{\partial t}=\frac{\pi d^{2}E^{2}}{2\hbar^{2}}\left|(\cos\theta)_{lm}\right|^{2}\delta(\omega_{0}-\omega)\,\,\,,\label{2}
\end{equation}
 where 
\begin{equation}
(\cos\theta)_{lm}=(\cos\theta)_{l+1,m;l,m}=-i\sqrt{\frac{(l+1)^{2}-m^{2}}{(2l+1)(2l+3)}}\,\,\,,\label{3}
\end{equation}
 $c_{lm}$ being the coefficients of the superposition of the wavefunctions
and $m$ being the quantum number of the component $L_{z}$ of the
angular momentum. The absorbed power (the spectrum) is 
\begin{equation}
\begin{array}{c}
P=\hbar\omega_{0}\sum_{m=-l}^{l}\frac{\partial\left|c_{lm}\right|^{2}}{\partial t}=\frac{\pi d^{2}E^{2}}{2\hbar}\omega_{0}\sum_{m=-l}^{l}\left|(\cos\theta)_{lm}\right|^{2}\delta(\omega_{0}-\omega)=\\
\\
=\frac{\pi d^{2}E^{2}}{6\hbar}\omega_{0}(l+1)\delta(\omega_{0}-\omega)=\frac{\pi d^{2}E^{2}}{6I}(l+1)^{2}\delta(\omega_{0}-\omega)
\end{array}\label{4}
\end{equation}
and the net absorbed power at finite temperatures is given by 
\begin{equation}
\begin{array}{c}
P_{th}=\frac{\pi d^{2}E^{2}}{2\hbar}\omega_{0}\times\\
\\
\times\sum_{m=-l}^{l}\left|(\cos\theta)_{lm}\right|^{2}\left[e^{-\beta\hbar^{2}l(l+1)/2I}-e^{-\beta\hbar^{2}(l+1)(l+2)/2I}\right]\delta(\omega_{0}-\omega)/Z\,\,\,,
\end{array}\label{5}
\end{equation}
 where 
\begin{equation}
Z=\sum_{l=0}(2l+1)e^{-\beta\hbar^{2}l(l+1)/2I}=\frac{2I}{\beta\hbar^{2}}\label{6}
\end{equation}
is the partition function and $\beta=1/T$ is the reciprocal of the
temperature $T$; we get 
\begin{equation}
\begin{array}{c}
P_{th}=\frac{\pi d^{2}E^{2}}{12I}(l+1)^{3}\left(\frac{\beta\hbar^{2}}{I}\right)^{2}e^{-\beta\hbar^{2}l(l+1)/2I}\delta(\omega_{0}-\omega)=\\
\\
=\frac{1}{2}P(l+1)\left(\frac{\beta\hbar^{2}}{I}\right)^{2}e^{-\beta\hbar^{2}l(l+1)/2I}\,\,.
\end{array}\label{7}
\end{equation}

For illustrative purposes we use $I=10^{-38}g\cdot cm^{2}$, which
is a typical numerical value for the molecular moment of inertia (molecular
mass $M=10^{5}$ electronic mass $m_{e}=10^{-27}g$ (heavy molecules),
the dipole length $r=10^{-8}cm$ ($1\textrm{\AA}$)), and get $\hbar/I=10^{11}s^{-1}\simeq1K$
($\omega_{0}=\hbar(l+1)/I$); at room temperature there are many levels
occupied, and we may use the inequality $\beta\hbar^{2}(l+1)/I\ll1$.

The classical dynamics corresponding to the hamiltonian given by equation
(\ref{1}) is governed by the equations of motion 
\begin{equation}
\ddot{\theta}=\dot{\varphi}^{2}\sin\theta\cos\theta\,\,,\,\, I\frac{d}{dt}(\dot{\varphi}\sin^{2}\theta)=0\,\,;\label{8}
\end{equation}
from the second equation (\ref{8}) we get $\dot{\varphi}=L_{z}/I\sin^{2}\theta$,
which indicates the conservation of the component $L_{z}$ of the
angular momentum (as it is well known, the angular momentum is conserved
in free rotations). The hamiltonian given by equation (\ref{1}) can
be written as 
\begin{equation}
H=\frac{1}{2}I\dot{\theta}^{2}+\frac{L_{z}^{2}}{2I\sin^{2}\theta}\,\,;\label{9}
\end{equation}
we can see that an effective potential function $U_{eff}=L_{z}^{2}/2I\sin^{2}\theta$
appears, which has a minimum for $\theta=\pi/2$. The motion may be
limited to small oscillations about the equatorial plane $\theta=\pi/2$.
Introducing $\delta\theta=\theta-\pi/2$ we get 
\begin{equation}
\frac{L_{z}^{2}}{2I\sin^{2}\theta}=\frac{L_{z}^{2}}{2I}+\frac{L_{z}^{2}}{2I}\delta\theta^{2}+...\label{10}
\end{equation}
 and 
\begin{equation}
H\simeq\frac{1}{2}I\delta\dot{\theta}^{2}+\frac{L_{z}^{2}}{2I}\delta\theta^{2}+\frac{L_{z}^{2}}{2I}\,\,.\label{11}
\end{equation}
We can see that there is a precession $\varphi=\omega_{0}t$ about
the $z$-axis and an oscillation $\delta\theta=A\cos(\omega_{0}t+\delta)$,
where $A$ is an undetermined amplitude and $\delta$ is an undetermined
phase, according to the small oscillations governed by the hamiltonian
given by equation (\ref{11}); the frequency $\omega_{0}$ is given
by $\omega_{0}=L_{z}/I$. We can check easily that the angular momentum
is conserved ($\dot{\mathbf{L}}=0$); the components of the angular
momentum are $L_{x}=IA\omega_{0}\cos\delta$, $L_{y}=IA\omega_{0}\sin\delta$,
and $L_{z}=I\omega_{0}$. We can rotate the equatorial plane $\theta=\pi/2$
by an angle given by $\sin\alpha=IA\omega_{0}/\sqrt{I^{2}\omega_{0}^{2}+I^{2}A^{2}\omega_{0}^{2}}\simeq A$,
such that the motion will be an in-plane motion.\cite{key-17,key-18}
This approximation corresponds to $L_{z}\simeq L$ ($m\simeq l$,
$L_{x}^{2}+L_{y}^{2}\ll L_{z}^{2}\simeq L^{2}$). 

The $\delta\theta$-motion governed by the harmonic-oscillator hamiltonian
given by equation (\ref{11}) can be quantized, the energy levels
being $\hbar\omega_{0}(n+1/2)$, $n=0,1,2...$, where $\omega_{0}=L_{z}/I=\hbar m/I$,
$m=0,1,2...$; the harmonic-oscillator frequency $\omega_{0}=\hbar m/I$
corresponds to the quantum-mechanical frequency $\omega_{0}=(E_{l+1}-E_{l})/\hbar=(\hbar/I)(l+1)$.
The interaction $H_{int}(t)=-dE\cos\theta\cos\omega t$, where $\theta=\pi/2+\delta\theta$
produces transitions of the type $n\rightarrow n+1$, with an absorbed
power 
\begin{equation}
P_{n}=\frac{\pi d^{2}E^{2}}{4I}(n+1)\delta(\omega_{0}-\omega)\label{12}
\end{equation}
(where the harmonic-oscillator matrix elements $(\delta\theta)_{n+1,n}=\sqrt{\hbar(n+1)/2I\omega_{0}}$
are used). The total power is obtained by summing $P_{n}$ with respect
to $n$ up to some value $N$ given by 
\begin{equation}
(\delta\theta)_{N+1,N}=\sqrt{\frac{\hbar(N+1)}{2I\omega_{0}}}=\sqrt{\frac{N+1}{2m}}\ll1\,\,\,,\label{13}
\end{equation}
 which gives 
\begin{equation}
P_{osc}=\sum_{n=0}^{N}P_{n}=\frac{\pi d^{2}E^{2}}{2I}m(m+1/2)\delta(\omega_{0}-\omega)\label{14}
\end{equation}
for $N=2m-1$; for large (and comparable) $m$ and $l$ this result
compares well with the exact absorbed power given by equation (\ref{4})
(up to a numerical factor $1/3$). We conclude that the separation
of the azimuthal and zenithal motions is a satisfactory harmonic-oscillator
approximation for the molecular spectra (for heavy molecules). (The
factor $1/3$ arises from the average $\overline{\cos^{2}\theta}=1/3$
of the factor $\cos^{2}\theta$ which apperas in equation (\ref{4})).

\textbf{High electric field.} Consider a constant, uniform electric
field $\mathbf{E}_{0}=E_{0}(0,0,1)$ oriented along the $z$-axis;
the potential energy of an electric dipole $\mathbf{d}=d(\sin\theta\cos\varphi,\sin\theta\sin\varphi,\cos\theta)$
of arbitrary orientation $\theta,\,\varphi$ is\\
 $U=-dE_{0}\cos\theta$. The hamiltonian of rotations in this field
is given by 
\begin{equation}
H=\frac{1}{2}I(\dot{\theta}^{2}+\dot{\varphi}^{2}\sin^{2}\theta)-dE_{0}\cos\theta\label{15}
\end{equation}
(for the Schroedinger equation with this hamiltonian see Refs. \cite{key-19,key-20}).
The component $L_{z}$ of the angular momentum is conserved, $\dot{\varphi}\sin^{2}\theta=L_{z}/I$;
consequently, an effective potential function 
\begin{equation}
U_{eff}=\frac{L_{z}^{2}}{2I\sin^{2}\theta}-dE_{0}\cos\theta\label{16}
\end{equation}
appears in the hamiltonian. We assume that the dipole energy $dE_{0}$
is much greater than the rotation energy $L_{z}^{2}/I$, which is
of the order of the temperature $T$. For typical value $d=10^{-18}statcoulomb\cdot cm$
and temperature $T=300K\simeq4\times10^{-14}erg$ this condition requires
an electric field $E_{0}\gg T/d=4\times10^{4}statvolt/cm\simeq1.2\times10^{9}V/m$.
This is a high electric field; for comparison, the electric field
created by an electron charge at distance $1\textrm{\AA}=10^{-8}cm$
is $4.8\times10^{-10}/10^{-16}=4.8\times10^{6}statvolt/cm$ (atomic
fields). Such a high electric field may appear as an internal field
in polar condensed matter (\emph{e.g.}, pyroelectrics, ferroelectrics).
At low temperatures the free molecular rotations may be hindered,
and the dipoles get quenched in parallel, equilibrium positions; they
may only perform small rotations and vibrations around these equilibrium
positions. The transitions from free rotations to small vibrations
around quenched positions in polar matter is seen in the curve of
the heat capacity \emph{vs} temperature.\cite{key-21,key-22} The
electric field produced by the nearest neighbours, averaged over their
small vibrations and rotations, gives rise to a local, static (mean)
electric field, which can be as high as the atomic fields. The condition
$E_{0}\gg T/d$ shows also that at lower temperatures (and high values
of the electric dipoles) the field $E_{0}$ may be weaker. Similarly,
high electric fields may appear locally near polar impurities with
large moments of inertia, embedded in polar matter. Under such conditions
the effective potential given by equation (\ref{16}) has a minimum
value for $\theta_{0}\simeq(L_{z}^{2}/IdE_{0})^{1/4}\simeq(T/dE_{0})^{1/4}\ll1$;
it can be expanded in powers of $\delta\theta=\theta-\theta_{0}$
around this minimum value, 
\begin{equation}
U_{eff}\simeq-dE_{0}+2dE_{0}\delta\theta^{2}\,\,;\label{17}
\end{equation}
the hamiltonian given by equation (\ref{15}) becomes 
\begin{equation}
H\simeq\frac{1}{2}I\delta\dot{\theta}^{2}+\frac{1}{2}I\omega_{0}^{2}\delta\theta^{2}-dE_{0}\,\,\,,\label{18}
\end{equation}
 where $\omega_{0}=2\sqrt{dE_{0}/I}$ is sometimes known as Rabi's
frequency;\cite{key-23,key-24} according to our condition of high
field, we have $\omega_{0}\gg10^{12}s^{-1}$ (we consider electric
fields that are not as high as to produce rotation frequencies comparable
with the molecular vibration frequencies). Therefore, the dipoles
exhibit quenched equilibrium positions in the static electric field
$E_{0}$, where they perform small oscillations and rotations. The
angle $\varphi$ rotates freely with the frequency $\dot{\varphi}\simeq L_{z}/I\sin^{2}\theta_{0}=\frac{1}{2}\omega_{0}$
($\varphi=\frac{1}{2}\omega_{0}t$). It is worth noting that the frequency
$\omega_{0}$ is determined by the external field $E_{0}$. An attempt
to derive the harmonic-oscillator hamiltonian given by equation (\ref{18})
has been made in Ref. \cite{key-25}.

Consider an external time-dependent field $\mathbf{E}(t)=E(t)(\sin\alpha,0,\cos\alpha)$,
$E(t)=E\cos\omega t$, which makes an angle $\alpha$ with the $z$-axis;
its interaction with the dipole is 
\begin{equation}
H_{int}=-dE(t)(\sin\alpha\sin\theta\cos\varphi+\cos\alpha\cos\theta)\,\,\,,\label{19}
\end{equation}
which provides two relevant interaction hamiltonians: 
\begin{equation}
\begin{array}{c}
H_{1int}=-\frac{1}{2}dE\sin\alpha\left[\cos(\omega+\frac{1}{2}\omega_{0})t+\cos(\omega-\frac{1}{2}\omega_{0})t\right]\delta\theta\,\,,\\
\\
H_{2int}=\frac{1}{2}dE\cos\alpha\cos\omega t\cdot\delta\theta^{2}\,\,.
\end{array}\label{20}
\end{equation}
 The interaction hamiltonian $H_{1int}$ produces transitions between
the harmonic-oscillator states $n$ and $n+1$ with the resonance
frequency $\Omega=\frac{1}{2}\omega_{0},\,\frac{3}{2}\omega_{0}$.
In general, for an interaction $H_{int}=h\cos\omega t$ (where $h$
is a time-independent interaction hamiltonian), the transition rate
between two states $n$ and $n+s$, with energies $E_{n}$, $E_{n+s}$
is 
\begin{equation}
\frac{\partial\left|c_{n+s,n}\right|^{2}}{\partial t}=\frac{\pi}{2\hbar^{2}}\left|h_{n+s,n}\right|^{2}\delta(\omega_{n;s}-\omega)\label{21}
\end{equation}
 in the first order of the perturbation theory, where $\omega_{n;s}=(E_{n+s}-E_{n})/\hbar$
and $c_{n+s,n}$ are the coefficients of the superposition of the
wavefunctions. For $H_{1int}$ we get 
\begin{equation}
\frac{\partial\left|c_{n+1,n}\right|^{2}}{\partial t}=\frac{\pi}{16\hbar I\omega_{0}}d^{2}E^{2}(n+1)\sin^{2}\alpha\delta(\omega-\Omega)\label{22}
\end{equation}
and the absorbed power 
\begin{equation}
\begin{array}{c}
P=\hbar\Omega\frac{\partial\left|c_{n+1,n}\right|^{2}}{\partial t}=\frac{\pi}{16I\omega_{0}}d^{2}E^{2}\Omega(n+1)\sin^{2}\alpha\delta(\omega-\Omega)=\\
\\
=\frac{\pi}{16I\omega_{0}}d^{2}E^{2}\Omega(n+1)\sin^{2}\alpha\delta(\omega-\Omega)\,\,.
\end{array}\label{23}
\end{equation}
In order to compute the mean power the thermal weigths $e^{-\beta\hbar\omega_{0}n}/\sum e^{-\beta\hbar\omega_{0}n}$
should be inserted, where $\beta=1/T$ is the inverse of the temperature
$T$; in addition, the reverse transitions must be taken into account.
Since $\beta\hbar\omega_{0}\gg1$, only the lowest states $n$ are
excited by interaction. The temperature dependence is given by 
\begin{equation}
\begin{array}{c}
P_{th}=\frac{\pi}{16I\omega_{0}}d^{2}E^{2}\Omega\sum_{n=0}(n+1)\left[e^{-\beta\hbar\omega_{0}n}-e^{-\beta\hbar\omega_{0}(n+1)}\right]\times\\
\\
\times\sin^{2}\alpha\delta(\omega-\Omega)/\sum_{n=0}e^{-\beta\hbar\omega_{0}n}\,\,\,,
\end{array}\label{24}
\end{equation}
 where the summation over $n$ is, in principle, limited. 

We should limit ourselevs to the lowest states of the harmonic oscillator,
since the oscillation amplitude $\delta\theta$ must be much smaller
than the angle $\theta_{0}$. The matrix element $(\delta\theta)_{n+1,n}=\sqrt{\hbar(n+1)/2I\omega_{0}}$
for the harmonic oscillator should be much smaller than $\theta_{0}\simeq(L_{z}^{2}/IdE_{0})^{1/4}$,
which implies $\hbar(n+1)\ll4L_{z}\simeq4\sqrt{IT}$; for typical
values $I=10^{-38}g\cdot cm^{2}$ we get $n\ll80$ for $T=300K$ (and
$n\ll8$ for $T=3K$). Consequently, for $\beta\hbar\omega_{0}\gg1$
we may extend the summation in equation (\ref{24}) to large values
of $n$; we get $P_{th}$ independent of temperature. Making use of
the expressions for the transverse components $L_{x,y}$ of the angular
momentum we get $L_{x}\simeq-(1/2)I\omega_{0}\theta_{0}\cos\omega_{o}t/2$
and $L_{y}\simeq-(1/2)I\omega_{0}\theta_{0}\sin\omega_{o}t/2$, which
show that the high-field approximation corresponds to $L_{x}^{2}+L_{y}^{2}\simeq L^{2}\gg L_{z}^{2}$
(small values of the component $L_{z}$).

Under the same conditions, the harmonic-oscillator hamiltonian given
by equation (\ref{18}) and the interaction hamiltonian $H_{2int}$
given by equation (\ref{20}), 
\begin{equation}
H^{'}=H+H_{2int}=\frac{1}{2}I\delta\dot{\theta}^{2}+\frac{1}{2}I\omega_{0}^{2}(1+h\cos\omega t)\delta\theta^{2}\,\,\,,\label{25}
\end{equation}
 where $h=\frac{E}{2E_{0}}\cos\alpha$, lead to the classical equation
of motion 
\begin{equation}
\delta\ddot{\theta}+\omega_{0}^{2}(1+h\cos\omega t)\delta\theta=0\,\,\,,\label{26}
\end{equation}
 which is the well-known equation of parametric resonance (Mathieu's
equation).\cite{key-26} As it is well known, beside periodic solutions,
the classical equation (\ref{26}) has also aperiodic solutions, which
may grow indefinitely with increasing time; these are (parametrically)
resonant solutions, which occur for $\omega$ in the neighbourhood
of $2\omega_{0}/n$, $n=1,2,3...$ . As we can see immediately, the
solutions of equation (\ref{26}) are determined by the initial conditions
$\delta\theta(t=0)$ and $\delta\dot{\theta}(t=0)$ (as for any second-order
differential equation). The initial conditions are vanishing due to
thermal fluctuations, so the classical solutions of equation (\ref{26})
are ineffective.

The quantum-mechanical dynamics is different. The interaction hamiltonian
$H_{2int}$ produces transitions between the harmonic-oscillator states
$n$ and $n+2$, due to the matrix elements of $\delta\theta^{2}$
(this is an example of a double-quanta process\cite{key-27}). These
transitions have frequency $2\omega_{0}$, in accordance with the
classical dynamics. The transition rate is 
\begin{equation}
\frac{\partial\left|c_{n+2,n}\right|^{2}}{\partial t}=\frac{\pi h^{2}}{128}\omega_{0}^{2}(n+1)(n+2)\delta(2\omega_{0}-\omega)\label{27}
\end{equation}
and the absorbed power 
\begin{equation}
\begin{array}{c}
P=2\hbar\omega_{0}\frac{\partial\left|c_{n+2,n}\right|^{2}}{\partial t}=\frac{\pi h^{2}}{64}\hbar\omega_{0}^{3}(n+1)(n+2)\delta(2\omega_{0}-\omega)\end{array}\label{28}
\end{equation}
where we may restrict, in principle, to the lowest states. The intensity
given by equation (\ref{28}) is small, because, especially, of the
factor $(E/E_{0})^{2}$ ($h=\frac{E}{2E_{0}}\cos\alpha$). The temperature
dependence is given by 
\begin{equation}
\begin{array}{c}
P_{th}=\frac{\pi h^{2}}{64}\hbar\omega_{0}^{3}\sum_{n=0}(n+1)(n+2)\times\\
\\
\times\left[e^{-\beta\hbar\omega_{0}(2n+1)}-e^{-\beta\hbar\omega_{0}(2n+3)}\right]\delta(2\omega_{0}-\omega)/\left[\sum_{n=0}e^{-\beta\hbar\omega_{0}n}\right]^{2}\,\,\,,
\end{array}\label{29}
\end{equation}
 in accordance with the direct transitions $n\rightarrow n+1\rightarrow n+2$
and the corresponding reverse transitions; $P_{th}$ is diminished
by the thermal factor $e^{-\beta\hbar\omega_{0}}$ for $\beta\hbar\omega_{0}\gg1$.

The parametric resonance disappears for $\alpha=\frac{\pi}{2}$, \emph{i.e.}
for the applied field $\mathbf{E}$ at right angle with the quenching
field $\mathbf{E}_{0}$. The effect of the parametric resonance depends
on the orientation of the (solid) sample; in amorphous samples the
average over angles $\alpha$ should be taken ($\overline{\cos^{2}\alpha}=\frac{1}{3}$).
In solids, the width of the absorption line (the damping parameter)
originates, very likely, in the dipolar interaction. Since the dipolar
interaction is taken mainly in the quenching effect, we may expect
a small damping, and, consequently, rather sharp resonance lines.
In liquids, beside the random distribution of the dipoles (and the
average over angle $\alpha)$, we may expect the usual motional narrowing
of the line. In gases the (internal) quenching field is weak, and
the parametric resonance is not likely to occur. 

\textbf{Weak electric field.} Consider now the opposite case, when
the field $E_{0}$ is weak, such that $dE_{0}\ll L_{z}^{2}/I$. The
effective potential $U_{eff}$ given by equation (\ref{16}) has a
minimum value for $\theta_{0}\simeq\frac{\pi}{2}$ and the hamiltonian
reduces to the free hamiltonian given by equation (\ref{11}); the
field $E_{0}$ brings only a small correction to the $\pi/2$-shift
in $\theta$, while its contribution to the hamiltonian is a second-order
effect. The angle $\varphi$ moves freely with angular velocity $\dot{\varphi}=\omega_{0}=L_{z}$/I.
In contrast with the high-field case, where the frequency $\dot{\varphi}$
is fixed by the high static field $E_{0}$, in the low-field case
we may quantize the $\varphi$-motion, according to $L_{z}=\hbar m$,
$m$ integer, such that $\omega_{0}=\frac{\hbar}{I}m$; the lowest
value of this frequency is $\hbar/I\simeq10^{11}s^{-1}$ for typical
values $I=10^{-38}g\cdot cm^{2}$. The molecular rotations are described
by a set of harmonic oscillators with frequencies $\omega_{0}=\frac{\hbar}{I}m$,
beside the $\varphi$-precession (which has the same frequencies $\omega_{0}$).
The energy quanta are $\hbar\omega_{0}=\frac{\hbar^{2}}{I}m$, with
the lowest value $\frac{\hbar^{2}}{I}=1K$ (for our numerical values).
The approximation described above is valid for $\delta\theta_{n+1,n}=\sqrt{\hbar(n+1)/2I\omega_{0}}\ll1$,
which leads to $\hbar(n+1)\ll2L_{z}$, or $n+1\ll m$. Similarly,
the transverse components of the angular momentum are very small,
$L_{x}^{2}+L_{y}^{2}\ll L^{2}\simeq L_{z}^{2}$ ($m\simeq l$); at
room temperature $m$ may acquire as high values as $m=300$. All
this is practically the same as for the free rotations. 

The interaction hamiltonian given by equation (\ref{19}) leads to
two relevant interactions 
\begin{equation}
\begin{array}{c}
H_{1int}=dE\cos\alpha\cos\omega t\cdot\delta\theta\,\,,\\
\\
H_{2int}=\frac{1}{4}dE\sin\alpha\left[\cos(\omega+\omega_{0})t+\cos(\omega-\omega_{0})t\right]\cdot\delta\theta^{2}\,\,.
\end{array}\label{30}
\end{equation}

The interaction $H_{1int}$ produces transitions between the harmonic-oscillator
states $n$ and $n+1$, with an absorbed power 
\begin{equation}
P_{n}=\frac{\pi}{4I}d^{2}E^{2}(n+1)\cos^{2}\alpha\delta(\omega_{0}-\omega)\,\,.\label{31}
\end{equation}

For $n\ll m$ we restrict ourselves to small values of $n$ in equation
(\ref{31}) and sum over a few values of $m$ in $\delta(\omega_{0}-\omega)=\delta(\hbar m/I-\omega)$
with the statistical weight $e^{-\beta\hbar^{2}m^{2}/2I}$ (low temperatures).
As long as $\hbar/I\gg\gamma$, where $\gamma$ is the resonance width,
the spectrum exhibits a few, distinct absorption lines at frequencies
$\omega_{0}=\hbar m/I$ (a band of absorption). In general, the temperature
dependence is given by 
\begin{equation}
\begin{array}{c}
P_{th}=\frac{\pi}{4I}d^{2}E^{2}\cos^{2}\alpha\cdot C\sum_{m>0}e^{-\beta\hbar^{2}m^{2}/2I}\times\\
\\
\times\left\{ \sum_{n=0}(n+1)\left[e^{-\beta\hbar\omega_{0}n}-e^{-\beta\hbar\omega_{0}(n+1)}\right]/\sum_{n=0}e^{-\beta\hbar\omega_{0}n}\right\} \delta(\omega_{0}-\omega)\,\,\,,
\end{array}\label{32}
\end{equation}
 where $\omega_{0}=\hbar m/I$ and $C\sum_{m>0}e^{-\beta\hbar^{2}m^{2}/2I}=1$.
At room temperature we may extend the summation over $n,\, m$ and
get the envelope of this function 
\begin{equation}
P_{th}=\frac{\pi}{4}d^{2}E^{2}\cos^{2}\alpha\sqrt{\frac{2\pi\beta}{I}}e^{-\beta I\omega^{2}/2}\,\,.\label{33}
\end{equation}

The interaction hamiltonian $H_{2int}$ given by equation (\ref{30})
produces transitions between states $n$ and $n+2$ (separated by
frequency $2\omega_{0}$) for external frequencies $\Omega=\omega_{0},\,3\omega_{0}$.
The absorbed power is 
\begin{equation}
P_{n}=\frac{\pi\hbar\Omega}{128I^{2}\omega_{0}^{2}}d^{2}E^{2}(n+1)(n+2)\sin^{2}\alpha\delta(\Omega-\omega)\,\,.\label{34}
\end{equation}
These parametric resonances, occurring at frequencies $\Omega=\omega_{0},\,3\omega_{0},$
are superposed over the transitions produced by $H_{1int}$. The temperature
dependence is given by
\begin{equation}
\begin{array}{c}
P_{th}=\frac{\pi\hbar}{128I^{2}}d^{2}E^{2}\sin^{2}\alpha\cdot C\sum_{m>0}\frac{\Omega}{\omega_{0}^{2}}e^{-\beta\hbar^{2}m^{2}/2I}\times\\
\\
\times\{\sum_{n=0}(n+1)(n+2)\left[e^{-\beta\hbar\omega_{0}(2n+1)}-e^{-\beta\hbar\omega_{0}(2n+3)}\right]/\\
\\
/\left[\sum_{n=0}e^{-\beta\hbar\omega_{0}n}\right]^{2}\}\delta(\Omega-\omega)\,\,;
\end{array}\label{35}
\end{equation}
 summation over $n$ gives 
\begin{equation}
P_{th}=\frac{\pi\hbar}{64I^{2}}d^{2}E^{2}\sin^{2}\alpha\cdot C\sum_{m>0}\frac{\Omega}{\omega_{0}^{2}}e^{-\beta\hbar^{2}m^{2}/2I}\frac{e^{-\beta\hbar\omega_{0}}}{(1+e^{-\beta\hbar\omega_{0}})^{2}}\delta(\Omega-\omega)\label{36}
\end{equation}
 whence we can get either the band of absorption or the envelope.

It is worth noting that the weak field $E_{0}$ does not appear explicitly
in the above formulae; its role is that of setting the $z$-axis,
to highlight the directional effect of the interaction field $E$
through the angle $\alpha$, and to reduce the conservation of the
angular momentum $\mathbf{L}$ to the conservation of only one component
($L_{z}$). In addition, the parametric resonances are a new feature
in the presence of the electric field. It is also worth noting that
the expansion of the effective potential function $U_{eff}$ in powers
of $\delta\theta$ is an approximation to free rotations with $L_{z}=const$,
instead of $\mathbf{L}=const$. 

It is also worth noting that a weak static electric field has an influence
on the statistical behaviour, as it is well known. Indeed, the hamiltonian
of rotations 
\begin{equation}
H=\frac{1}{2}I(\dot{\theta}^{2}+\dot{\varphi}^{2}\sin^{2}\theta)\label{37}
\end{equation}
 can also be written as 
\begin{equation}
H=\frac{1}{2I}P_{\theta}^{2}+\frac{1}{2I\sin^{2}\theta}P_{\varphi}^{2}\label{38}
\end{equation}
with the (angular) momenta $P_{\theta}=I\dot{\theta}$ and $P_{\varphi}=I\dot{\varphi}\sin^{2}\theta$.
The classical statistical distribution is 
\begin{equation}
const\cdot dP_{\theta}dP_{\varphi}d\theta e^{-\beta P_{\theta}^{2}/2I}e^{-\beta P_{\varphi}^{2}/2I\sin^{2}\theta}\,\,\,,\label{39}
\end{equation}
 or, integrating over momenta, $\frac{1}{2}\sin\theta d\theta$. In
the presence of the field we have the distribution $\simeq\frac{1}{2}\sin\theta d\theta\cdot e^{\beta\mathbf{d}\mathbf{E}_{0}}$
(since $\beta dE_{0}\ll1$), which leads, for example, to $\overline{\cos\theta}=\beta dE_{0}/3$.
This is the well-known Curie-Langevin-Debye law.\cite{key-28}-\cite{key-31}
In the quantum-mechanical regime, for $dE_{0}\ll\hbar^{2}/I$, the
interaction $-dE_{0}\cos\theta$ brings a second-order contribution
to the energy levels $E_{l}=\hbar^{2}l(l+1)/2I$ and renormalize the
wavefunction in the first-order of the theory of perturbation; using
these renormalized wavefunctions, there appear diagonal matrix elements
of $\cos\theta$, which we denote by $(\widetilde{\cos\theta})_{lm,lm}$;
the mean value of this quantity is given by $\overline{\cos\theta}=\sum\widetilde{(\cos\theta)}_{lm,lm}\Delta(\beta E_{l})e^{-\beta E_{l}}/\sum e^{-\beta E_{l}}$,
which leads to the classical result $=\beta dE_{0}/3$, as expected.

\textbf{Dipolar interaction.} Although many molecules possess an electric
dipole moment $d$, even in their ground state, usually the dipole-dipole
interaction is neglected in rarefied condensed matter, on the ground
that the distance between the dipoles is large. In these conditions,
at finite temperatures, the electric dipoles are randomly distributed;
they get slightly aligned in the presence of a static external electric
field $\mathbf{E}_{0}$, which provides a small interaction energy,
leading to an induced orientational polarization $\overline{d}=\beta d^{2}E_{0}/3$,
as noted above. 

For typical values of the dipole moments $d=10^{-18}statcoulomb\cdot cm$
separated by distance of the order $a=10^{-8}cm$ ($1\textrm{\AA}$)
the interaction energy is $\simeq d^{2}/a^{3}=10^{-12}erg\simeq10^{3}K$
($1eV=1.6\times10^{-12}erg$, $1K=1.38\times10^{-16}erg$, $1eV=1.1\times10^{4}K$).
This is not a small energy (it corresponds approximately to a frequency
$10^{13}Hz$), and, apart from special circumstances, the electric
dipole-dipole interaction cannot be neglected. (The estimation given
here should take into account the time average of the dipole interaction
energy with respect to molecular motion). The corresponding dipolar
field is of the order $d/a^{3}=10^{6}statvolt/cm$ (\emph{i.e.}, of
the order of the atomic fields).

The interaction energy of two dipoles $\mathbf{d}_{1}$ and $\mathbf{d}_{2}$
separated by distance $\mathbf{a}$ is given by 
\begin{equation}
U=-\frac{3(\mathbf{d}_{1}\mathbf{d}_{2})a^{2}-(\mathbf{d}_{1}\mathbf{a})(\mathbf{d}_{2}\mathbf{a})}{a^{5}}\,\,.\label{40}
\end{equation}
We introduce the angles $(\theta_{1},\,\varphi_{1})$ and $(\theta_{2},\,\varphi_{2})$
for the directions of the two dipoles with respect to the axis $\mathbf{a}$
and the interaction energy becomes 
\begin{equation}
U=-\frac{d_{1}d_{2}}{a^{3}}[2\cos\theta_{1}\cos\theta_{2}+3\sin\theta_{1}\sin\theta_{2}\cos(\varphi_{1}-\varphi_{2})]\,\,;\label{41}
\end{equation}
this energy has four extrema for $\theta_{1}=\theta_{2}=0,\,\pi/2$
and $\varphi_{1}-\varphi_{2}=0,\,\pi$; only for $\theta_{1}=\theta_{2}=\pi/2$,
$\varphi_{1}-\varphi_{2}=0$ the interaction energy has a local minimum;
in the neighbourhood of this minimum value the interaction energy
behaves like
\begin{equation}
\begin{array}{c}
U=\frac{d_{1}d_{2}}{a^{3}}[-3+\frac{3}{2}(\delta\theta_{1}^{2}+\delta\theta_{2}^{2})-2\delta\theta_{1}\delta\theta_{2}+\frac{3}{2}(\delta\varphi_{1}-\delta\varphi_{2})^{2}]=\\
\\
=\frac{d_{1}d_{2}}{a^{3}}[-3+\frac{1}{4}(\delta\theta_{1}+\delta\theta_{2})^{2}+\frac{5}{4}(\delta\theta_{1}-\delta\theta_{2})^{2}+\frac{3}{2}(\delta\varphi_{1}-\delta\varphi_{2})^{2}]\,\,\,,
\end{array}\label{42}
\end{equation}
 where $\delta\theta_{1,2}=\theta_{1,2}-\pi/2$ are small deviations
of the angles $\theta_{1,2}$ from the polarization axis $\pi/2$;
similarly, $\delta\varphi_{1,2}$ are small deviations of the angles
$\varphi_{1,2}$ from their equilibrium values $\varphi_{1,2}$, subjected
to the condition $\varphi_{1}-\varphi_{2}=0$. It follows that the
electric dipoles exhibit quenched equilibrium positions $\theta_{1}=\theta_{2}=\pi/2$,
$\varphi_{1}-\varphi_{2}=0$, such that they are parallel to each
other and perpendicular to the distance between them; they may perform
small rotations and vibrations around these equilibrium positions.
For the other three extrema the interaction energy has either a saddle
point ($\theta_{1}=\theta_{2}=0$, $\varphi_{1}-\varphi_{2}=0,\,\pi$)
or a maximum ($\theta_{1}=\theta_{2}=\pi/2$, $\varphi_{1}-\varphi_{2}=\pi$).
It is very likely that the structural environment is distorted such
as the dipoles take advantage of the energy minimum. For instance,
a structural elongation along the direction $\theta_{1}=\theta_{2}=0$
decreases appreciably the dipolar interaction along this direction
(which goes like $1/a^{3}$!), such that the corresponding contribution
to the energy may be neglected. Under such circumstances, for not
too high temperatures, we may expect the dipoles to be (spontaneously)
aligned along an arbitrary axis (in isotropic matter), giving rise
to an electric (macroscopic) polarization along such an axis. The
neglect of the interaction along the direction $\theta_{1}=\theta_{2}=0$
makes this model highly anisotropic, with a layered structure of the
aligned dipoles. 

These considerations are based on the dipolar interaction given by
equation (\ref{40}), which, in principle, is valid for distances
$a$ much longer than the dimension of the dipoles. However, since
the dipolar interaction decreases rapidly with increasing distance,
we may also use it for distances equal to a few dipole lengths. In
addition, for heavy molecules the charge imbalance implies a large
charge and, consequently, a small displacement, which amounts to a
more localized dipole; so, the condition for validity of the dipolar
interaction may be fulfilled much more satisfactorily than we use
to think. It is relevant in this respect the analysis made in Ref.
\cite{key-32}. 

As it is well known, pyroelectrics (or electrets) have a permanent
electric polarization;\cite{key-33} if the polarization is singular
just below a critical temperature and vanishes above, those substances
are called ferroelectrics (in the state above the critical temperature
they are also called paraelectrics); it seems that all these substances
are piezoelectric. There are also structural modifications associated
with finite discontinuities in polarization, a typical example being
barium titanate ($BaTiO_{3})$; the dimension of the elementary cell
in the crystal of $BaTiO_{3}$ is $a\simeq4\times10^{-8}cm$ ($4\textrm{\AA}$);
the dipole of a cell is $d\simeq5\times10^{-18}statcoulomb\cdot cm$
(the saturation polarization - the dipole moment per unit volume -
at room temperature is $8\times10^{4}statcoulomb\cdot cm$); if $Ba^{2+}$
and $Ti^{4+}$ are displaced by $\delta$ with respect to $O^{2-}$,
then the dipole moment $d$ is achieved for a slight displacement
$\delta=0.1\textrm{\AA}$; we can see that the distance $a$ between
the dipoles is much longer than the dimension $\delta$ of the dipoles.
In addition, $BaTiO_{3}$ exhibits several structural modifications
(from cubic to tetragonal to monoclinic to rhombohedral with decreasing
temperature), in all polarized phases the structure being elongated
along the direction of polarization.\cite{key-34} 

In a continuum model of polarized substance the dipolar interaction
given by equation (\ref{42}) (with identical dipoles $d$) gives
the interaction hamiltonian 
\begin{equation}
H_{int}=\frac{1}{a^{3}}\int d\mathbf{r}\left[\frac{d^{2}}{a^{3}}\delta\theta^{2}+\frac{5d^{2}}{4a}(grad\delta\theta)^{2}+\frac{3d^{2}}{2a}(grad\delta\varphi)^{2}\right]\,\,\,,\label{43}
\end{equation}
 which, together with the kinetic part, leads to the full hamiltonian
\begin{equation}
\begin{array}{c}
H=\frac{1}{a^{3}}\int d\mathbf{r}[\frac{1}{2}I\dot{\delta\theta}^{2}+\frac{1}{2}I\dot{\delta\varphi}^{2}+\frac{1}{2}I\omega_{0}^{2}\delta\theta^{2}+\\
\\
+\frac{1}{2}Iv_{\theta}^{2}(grad\delta\theta)^{2}+\frac{1}{2}Iv_{\varphi}^{2}(grad\delta\varphi)^{2}\,\,\,,
\end{array}\label{44}
\end{equation}
 where $I$ is the moment of inertia of the dipoles and $\omega_{0}^{2}=2d^{2}/Ia^{3}$,
$v_{\theta}^{2}=5d^{2}/2Ia=5\omega_{0}^{2}a^{2}/4$, $v_{\varphi}^{2}=3d^{2}/Ia=3\omega_{0}^{2}a^{2}/2$.
The dipole density $1/a^{3}$ should include the number of nearest
neighbours; if we restrict ourselves to the highly anisotropic (layered)
model, then the hamiltonian density in equation (\ref{44}) is two-dimensional.
We can see that the dipolar interaction may generate dipolar waves
(waves of orientational polarizability), governed by the wave equations
\begin{equation}
\frac{d^{2}}{dt^{2}}\delta\theta+\omega_{0}^{2}\delta\theta-v_{\theta}^{2}\Delta\delta\theta=0\,\,,\,\,\frac{d^{2}}{dt^{2}}\delta\varphi-v_{\varphi}^{2}\Delta\delta\varphi=0\,\,;\label{45}
\end{equation}
 the spectrum of these dipolar waves is $\omega_{\theta}^{2}=\omega_{0}^{2}+v_{\theta}^{2}k^{2}$
and, respectively, $\omega_{\varphi}^{2}=v_{\varphi}^{2}k^{2}$ (in
the layered model the wavevector $\mathbf{k}$ is two-dimensional);
for typical values $d=10^{-18}statcoulomb\cdot cm$, $a=10^{-8}cm$
and $I=10^{-38}g\cdot cm^{2}$ we get the frequency $\omega_{0}\simeq10^{13}s^{-1}$
(infrared region) and the wave velocities $v_{\theta,\varphi}\simeq10^{5}cm/s$
(the wavelengths are $\lambda_{\theta,\varphi}\simeq\pi\sqrt{5}a,\,\pi\sqrt{6}a$).
It is worth noting that the coordinates $\delta\theta$, $\delta\varphi$
are the tilting angles of the polarization with respect to its equilibrium
direction. Tentatively, we may call these polar-matter modes \textquotedbl{}dipolons\textquotedbl{}.
They contribute to the anomalous heat-capacity curve \emph{vs} temperature. 

The dipolar waves can couple to an external time-dependent electric
field. Let $\mathbf{E}(\mathbf{r},t)=\mathbf{E}\cos(\omega t-\mathbf{kr})$
be a radiation electric field (plane wave) which makes an angle $\alpha$
with the polarization direction; the interaction hamiltonian is 
\begin{equation}
H^{'}=-\frac{1}{a^{3}}\int d\mathbf{r}(\mathbf{d}\mathbf{E})\mathbf{\cos}(\mathbf{\omega}t\mathbf{-\mathbf{kr}})\mathbf{\,\,,}\label{46}
\end{equation}
 where $\mathbf{E}=E(\sin\alpha\cos\varphi^{'},\,\sin\alpha\sin\varphi^{'},\,\cos\alpha)$
\\
and $\mathbf{d}=d(\sin\delta\theta\cos\varphi$,~$\sin\delta\theta\sin\varphi,\,\cos\delta\theta$);
we may limit ourselves to $\varphi=\varphi^{'}$, and get 
\begin{equation}
H^{'}=-\frac{1}{a^{3}}\int d\mathbf{r}(dE)(\delta\theta\sin\alpha-\frac{1}{2}\delta\theta^{2}\cos\alpha)\mathbf{\cos(\omega}t\mathbf{-\mathbf{kr}})\label{47}
\end{equation}
(up to irrelevant terms); we can see that the $\varphi$-waves do
not couple to the external electric field (within the present approximation).
Moreover, since the wavelength of the radiation field is much longer
than the wavelength of the dipolar interaction ($v_{\theta,\varphi}\ll c$,
where $c$ is the speed of light), we may drop out the spatial dependence
(spatial dispersion) both in equation (\ref{45}) and in the interaction
hamiltonian $H^{'}$; we are left with the equation of motion of a
harmonic oscillator under the action of an external force, 
\begin{equation}
\frac{d^{2}}{dt^{2}}\delta\theta+\omega_{0}^{2}\delta\theta=\frac{dE}{I}\sin\alpha\cos\omega t-\frac{dE}{I}\delta\theta\cos\alpha\cos\omega t\,\,.\label{48}
\end{equation}
The first interaction term gives 
\begin{equation}
\frac{d^{2}}{dt^{2}}\delta\theta+\omega_{0}^{2}\delta\theta_{1}+2\gamma\frac{d}{dt}\delta\theta=\frac{dE}{I}\sin\alpha\cos\omega t\,\,\,,\label{49}
\end{equation}
 where a damping term ($\gamma$ coefficient) has been introduced;
this is the equation of motion of a harmonic oscillator under the
action of a harmonic force; the (particular) solution is 
\begin{equation}
\delta\theta_{1}=a\cos\omega t+b\sin\omega t\,\,\,,\label{50}
\end{equation}
where 
\begin{equation}
a=-\frac{dE}{2I\omega_{0}}\sin\alpha\frac{\omega-\omega_{0}}{(\omega-\omega_{0})^{2}+\gamma^{2}}\,\,,\,\, b=\frac{dE}{2I\omega_{0}}\sin\alpha\frac{\gamma}{(\omega-\omega_{0})^{2}+\gamma^{2}}\label{51}
\end{equation}
for $\omega$ near $\omega_{0}$; we get a resonance for $\omega=\omega_{0}$;
the absorbed mean power is 
\begin{equation}
P=dE\sin\alpha\overline{\cos\omega t\dot{\delta\theta}_{1}}=\frac{1}{2}dE\sin\alpha\cdot b\omega_{0}=\frac{\pi}{4I}d^{2}E^{2}\sin^{2}\alpha\delta(\omega_{0}-\omega)\,\,.\label{52}
\end{equation}

The second interaction term in equation (\ref{48}) gives the Mathieu's
equation 
\begin{equation}
\ddot{\delta\theta}_{2}+\omega_{0}^{2}(1+h\cos\omega t)\delta\theta_{2}=0\,\,\,,\label{53}
\end{equation}
 where $h=(dE/I\omega_{0}^{2})\cos\alpha$ (a damping term can be
included). As it was discussed before the thermal fluctuations wipe
out the parametric resonances associated with this equation. All the
above considerations are valid for a classical dynamics. The quantization
of the hamiltonians $H$ and $H^{'}$ given by equations (\ref{44})
and (\ref{47}) (which is performed according to the well-known standard
rules), leads to standard absorption and emission processes, and to
quantum transitions similar with equations (\ref{27})-(\ref{29}).
It is worth noting that the static electric field $E_{0}$ in equations
(\ref{27})-(\ref{29}) is replaced here by $E_{0}=d/2a^{3}$ (by
comparing the frequencies $\omega_{0}$ given in equations (\ref{18})
and, respectively, (\ref{44})), as expected for a (high) electric
field generated by a dipolar interaction.

The spontaneous polarization caused by the dipolar interaction as
described above may appear in polarization domains, randomly distributed
in polar matter (pyroelectrics, ferroelectrics), or in granular matter,
where charges may accumulate at the interfaces.\cite{key-35}-\cite{key-41}
This is known as the Maxwell-Wagner-Sillars effect (an average over
the angle $\alpha$ should then be taken in the absorbed power). In
the latter case the distance between the dipoles is much larger than
the atomic distances and, consequently, the characteristic frequency
$\omega_{0}$ is much lower; for instance, for a distance $a=1\mu m$
($10^{4}\textrm{\AA})$ we get a frequency $\omega_{0}\simeq10MHz$. 

\textbf{Highly-oscillating electric fields.} High-power lasers may
provide strong electric fields which oscillate in time with a frequency
$\omega_{h}$ much higher than the frequencies of molecular rotations
or vibrations. Usually, the frequency $\omega_{h}$ is in the optical
range, $\omega_{h}=2\pi\times10^{15}s^{-1}$, and the strength of
the electric field may attain values as high as $E_{0}=10^{9}statvolt/cm$
for laser intensities $10^{20}W/cm^{2}$. Under the action of such
strong fields the molecules are usually ionized, but the molecular
ions retain their electric dipoles which perform a non-relativistic
motion. (Indeed, the non-relativistic approximation is ensured by
the inequality $\eta=qA_{0}/2Mc^{2}\ll1$, where $q$ is the charge
of the particle with masss $M$ and $A_{0}$ is the amplitude of the
vector potential; for a proton in a potential $A_{0}=5\times10^{3}statvolt$,
corresponding to the field amplitude $E_{0}=10^{9}statvolt/cm,$ we
get $\eta=10^{-3}\ll1$).

Consider an electric field $E_{0}\cos\omega_{h}t$ oriented along
the $z$-axis. An electric dipole $d$ acted by this field performs
rapid oscillations of an angle $\alpha$ about, in general, a certain
angle $\theta$ measured with respect to the $z$-axis; the angle
$\theta$ may perform slow oscillations; we assume $\alpha\ll\theta$.
The equation of motion can be written as 
\begin{equation}
I\ddot{\alpha}=-dE_{0}\sin(\theta+\alpha)\cos\omega_{h}t\simeq-dE_{0}\sin\theta\cos\omega_{h}t\,\,;\label{54}
\end{equation}
 the corresponding kinetic energy is $E_{kin}=I\dot{\alpha}^{2}/2=(d^{2}E_{0}^{2}/2I\omega_{h}^{2})\sin^{2}\theta\sin^{2}\omega_{h}t$;
its time average
\begin{equation}
\overline{E}_{kin}=\frac{d^{2}E_{0}^{2}}{4I\omega_{h}^{2}}\sin^{2}\theta\label{55}
\end{equation}
 replaces the interaction energy $-dE_{0}\cos\theta$ of the static
field in the effective potential energy $U_{eff}$ given by equation
(\ref{16}); the effective potential becomes 
\begin{equation}
U_{eff}=\frac{L_{z}^{2}}{2I\sin^{2}\theta}+\frac{d^{2}E_{0}^{2}}{4I\omega_{h}^{2}}\sin^{2}\theta\,\,.\label{56}
\end{equation}
This function has a minimum value for $\widetilde{\theta}_{0}=\arcsin\theta_{0}/R^{1/4}$
and $\widetilde{\theta^{'}}_{0}=\pi-\widetilde{\theta}_{0}$, where
$R=dE_{0}/2I\omega_{h}^{2}$ is a renormalization factor and $\theta_{0}=(L_{z}^{2}/IdE_{0})^{1/4}<R^{1/4}$,
\emph{i.e.} the $\theta_{0}$ value corresponding to high fields,
as expected; it is worth noting that there are two values of the equilibrium
angle: $\widetilde{\theta}_{0}$ and $\pi-\widetilde{\theta}_{0}$.
The dipole may perform small vibrations about these equilibrium angles
with the frequency $\widetilde{\omega}_{0}=\omega_{0}\sqrt{3R/4}$,
where $\omega_{0}=2\sqrt{dE_{0}/I}$ is the characteristic frequency
for high static fields given in equation (\ref{18}) (for$\widetilde{\theta}_{0}\ll1$).
We can see that for highly-oscillating electric fields we get the
results for static fields renormalized according to $E_{0}\rightarrow\widetilde{E}_{0}=E_{0}R$.

From $\theta_{0}/R^{1/4}<1$ and $\alpha=(dE_{0}/I\omega_{h}^{2})\widetilde{\theta}_{0}\ll\widetilde{\theta}_{0}$
we get the inequalities 
\begin{equation}
\frac{L_{z}^{2}}{IdE_{0}}<\frac{dE_{0}}{2I\omega_{h}^{2}}\ll1\label{57}
\end{equation}
 (which are compatible because $L_{z}\ll I\omega_{h})$; these inequalities
imply 
\begin{equation}
\frac{\sqrt{2}L_{z}\omega_{h}}{d}<E_{0}\ll\frac{2I\omega_{h}^{2}}{d}\,\,.\label{58}
\end{equation}
For $L_{z}^{2}/I=T$ and our numerical parameters $I=10^{-38}g\cdot cm^{2}$,
$T=300K=4\times10^{-14}erg$, $d=10^{-18}statcoulomb\cdot cm$ and
$\omega_{h}=2\pi\cdot10^{15}s^{-1}$ we get approximately $10^{8}statvolt/cm<E_{0}\ll10^{10}statvolt/cm$,
which corresponds to a renormalization parameter $R=10^{-10}E_{0}/8\pi^{2}\ll1$.
We conclude that in strong, highly-oscillating electric fields, like
those provided by high-power lasers, the molecular rotation spectra
are affected in the same manner as in static electric field, providing
the time-dependent field strength is renormalized by the factor $R\ll1$
introduced here. It is worth noting that the interaction $-dE_{0}\cos\theta\cos\omega_{h}t$
linear in the field is replaced by an effective interaction which
is quadratic in the field, as shown in equation (\ref{56}); while
this effective interaction affects the slow rotations, it does not
couple to the (slow) translation motion. 

\textbf{Discussion and conclusions.} We have shown here that the rotations
of a heavy molecule (viewed as a spherical pendulum) can be approximated
by azimuthal rotations and zenithal oscillations. The molecular electric-dipole
moment couples to a time-dependent external electric field and gives
rise to rotation (and vibration) molecular spectra. Arguments have
been given that in polar matter there could appear local, strong,
static electric fields, which can lead to quenched equilibrium positions
for the dipoles and a macroscopic electric polarization. The small
rotations and oscillations which these dipoles may perform about their
equilibrium positions give rise to special features in the spectrum,
in particular to parametric resonances. Similar parametric resonances
appear in the presence of weak static electric fields. It was shown
that the dipole-dipole interaction can lead to an equilibrium state
of quenched dipoles, which possesses a macroscopic polarization; the
motion of this macroscopic polarization proceeds by particular modes
which have been tentatively called \textquotedbl{}dipolons\textquotedbl{}
(polarization waves). The excitation of these modes may also lead
to parametric resonances. It was also shown that a strong, highly-oscillating
electric field, like the fields provided by the high-power lasers,
behaves in the same manner as static electric fields, provided they
are renormalized by factors much smaller than unity (factor $R$ above).

All the discussion made in this paper for electric dipole moments
can also be applied to magnetic moments, magnetic fields, magnetization
and magnetic matter (\emph{e.g.}\textbf{\emph{,}} ferromagnetics).
The main difference is the magnitude; the nuclear magnetic moments
are five orders of magnitude smaller than the electric dipole moments
($\mu\simeq10^{-23}erg/Gs$); if the magnetic moments are in thermal
equilibrium, their interaction energy $\mu^{2}/a^{3}\simeq10^{-6}K$
is effective at much lower temperatures; the characteristic frequency
of \textquotedbl{}electric dipolons\textquotedbl{} $\omega_{0}=\sqrt{2d^{2}/Ia^{3}}\simeq10^{13}s^{-1}$
becomes $\omega_{0}=\sqrt{2\mu^{2}/Ia^{3}}\simeq10^{8}s^{-1}$ for
\textquotedbl{}magnetic dipolons\textquotedbl{}. For electronic magnetic
moments $\mu\simeq10^{-20}erg/Gs$ the interaction energy is $\simeq1K$
and the characteristic frequency is $\omega_{0}\simeq10^{11}s^{-1}$.
If the magnetic moment is higher by a factor of, say, $5$ and the
number of nearest neighbours is $4$, then the effective magnetic
dipolar energy (for electronic moments) increases to $\simeq100K$,
which is of the order of magnitude of usual ferromagnetic transition
temperatures; then, the \textquotedbl{}magnetic dipolons\textquotedbl{}
become magnons (ferromagnetic resonances).\cite{key-34} The dipole
interaction as source of ferromagnetism is different from the Weiss
mean field approach; it resembles more the Bloch theory of magnons.\cite{key-42} 

\textbf{Acknowledgments.} The authors are indebted to the members
of the Laboratory of Theoretical Physics at Magurele-Bucharest for
useful discussions. This work was supported by UEFISCDI Grants Core
Program \#09370108-2009/Ph12/2014/ELI-NP of the Romanian Governmental
Agency for Research.


\begin{thebibliography}{10}
\bibitem[1]{key-1}E. Wrede, \textquotedbl{}Uber die Ablenkung von
Molekularstrahlen elektrischer Dipolemolekule im inhomogenen elektrischen
Feld\textquotedbl{}, Z. Phys. \textbf{44} 261-268 (1927).

\bibitem[2]{key-2}D. H. Parker and R. B. Bernstein, \textquotedbl{}Oriented
molecule beams via the electrostatic hexapole: preparation, characterization,
and reactive scattering\textquotedbl{}, Ann. Rev. Phys. Chem. \textbf{40}
561-595 (1989).

\bibitem[3]{key-3}H. J. Loesch, \textquotedbl{}Orientation and alignment
in reactive beam collisions: recent progress\textquotedbl{}, Ann.
Rev. Phys. Chem. \textbf{46} 555-594 (1995).

\bibitem[4]{key-4}H. L. Bethlem, G. Berden, F. M. H. Crompvoets,
R. T. Jongma, A. J. A. van Roij and G. Meijer, \textquotedbl{}Electrostatic
trapping of amonia molecules\textquotedbl{}, Nature \textbf{406} 491-494
(2000).

\bibitem[5]{key-5}P. Dogourd, I. Compagnon, F. Lepine, R. Antoine,
D. Rayane and M. Broyer, \textquotedbl{}Beam deviation of large polar
molecules in static electric fields: theory and experiment\textquotedbl{},
Chem. Phys. Lett. \textbf{336} 511-517 (2001).

\bibitem[6]{key-6}Y. Yamakita, S. R. Procter, A. L. Goodgame, T.
P. Softley and F. Merkt, \textquotedbl{}Deflection and deceleration
of hydrogen Rydberg molecules in inhomogeneous electric fields\textquotedbl{},
J. Chem. Phys. \textbf{121} 1419-1431 (2004).

\bibitem[7]{key-7}M. Abd El Rahim, R. Antoine, M. Broyer, D. Rayane
and P. Dogourd, \textquotedbl{}Asymmetric top rotors in electric fields:
influence of chaos and collisions in molecular beam deflection experiments\textquotedbl{},
J. Phys. Chem. \textbf{A109} 8507-8514 (2005).

\bibitem[8]{key-8}B. Friedrich and D. R. Herschbach, \textquotedbl{}Polarization
of molecules induced by intense nonresonant laser fields\textquotedbl{},
J. Phys. Chem. \textbf{99} 15686-15693 (1995).

\bibitem[9]{key-9}B. Friedrich and D. R. Herschbach, \textquotedbl{}Enhanced
orientation of polar molecules by combined electrostatic and nonresonant
induced dipole forces\textquotedbl{}, J. Chem. Phys. \textbf{111}
6157-6160 (1999).

\bibitem[10]{key-10}B. Friedrich and D. R. Herschbach, \textquotedbl{}Manipulating
molecules via combined static and laser fields\textquotedbl{}, J.
Phys. Chem. \textbf{A103} 10280-10288 (1999).

\bibitem[11]{key-11}L. Cai, J. Marango and B. Friedrich, \textquotedbl{}Time-dependent
alignment and orientation of molecules in combined electrostatic and
pulsed nonresonant laser fields\textquotedbl{}, Phys. Rev. Lett. \textbf{86}
775-778 (2001).

\bibitem[12]{key-12}C. A Arango, W. W. Kennerly and G. S. Ezra, \textquotedbl{}Classical
and quantum mechanics of diatomic molecules in tilted fields\textquotedbl{},
J. Chem Phys \textbf{122} 184303 (2005)(1-15).

\bibitem[13]{key-13}J. P. Salas, \textquotedbl{}Classical dynamics
of polar diatomic molecules in external fields\textquotedbl{}, Eur.
Phys. J. \textbf{D41} 95-102 (2007).

\bibitem[14]{key-14}M. Lemeshko, R. V. Krems, J. M. Doyle and S.
Kais, \textquotedbl{}Manipulation of molecules with electromagnetic
fields\textquotedbl{}, Mol. Phys. \textbf{111} 1648-1682 (2013).

\bibitem[15]{key-15}R. N. Zare, \emph{Angular Momenta}, Wiley-Interscience,
NY (1988).

\bibitem[16]{key-16}G. Herzberg, \emph{Molecular Spectra and Molecular
Structure}, vol.1, Van Nostrand, Princeton (1950).

\bibitem[17]{key-17}A. Sommerfeld, \emph{Vorlesungen uber Theoretische
Physik}, Bd.1, \emph{Mechanik}, Akad. Verlagsgeselschaft, Lepzig (1968).

\bibitem[18]{key-18}L. Landau and E. Lifshitz, \emph{Course of Theoretical
Physics}, vol.1, \emph{Mechanics}, Elsevier, Oxford (1976).

\bibitem[19]{key-19}N. Froman, P. O. Froman and K. Larssen, \textquotedbl{}Rotation
of a rigid diatomic molecule in a homogeneous electric field. I. Schrodinger
equation. Quantization conditions according to phase integral method\textquotedbl{},
Phil. Trans. Roy. Soc. Lond. \textbf{A347} 1-22 (1994).

\bibitem[20]{key-20}P. O. Froman and K. Larssen, \textquotedbl{}Rotation
of a rigid diatomic molecule in a homogeneous electric field. II.
Energy levels when the field iz zero, very weak or very strong\textquotedbl{},
Phil. Trans. Roy. Soc. Lond. \textbf{A347} 23-35 (1994).

\bibitem[21]{key-21}L. Pauling, \textquotedbl{}The rotational motion
of molecules in crystals\textquotedbl{}, Phys. Rev. \textbf{36} 430
(1930).

\bibitem[22]{key-22}T. E. Stern, \textquotedbl{}The symmetrical spherical
oscillator, and the rotational motion of homopolar molecules in crystals\textquotedbl{},
Proc. Roy. Soc. \textbf{A130} 551 (1931).

\bibitem[23]{key-23}I. I. Rabi, \textquotedbl{}On the process of
space quantization\textquotedbl{}, Phys. Rev. \textbf{49} 324 (1936).

\bibitem[24]{key-24}I. I. Rabi, \textquotedbl{}Space quantization
in a gyrating magnetic field\textquotedbl{}, Phys. Rev. \textbf{51}
652 (1937).

\bibitem[25]{key-25}K. von Meyenn, \textquotedbl{}Rotation von zweiatomigen
Dipolmolekulen in starken elektrischen Feldern\textquotedbl{}, Z.
Phys. \textbf{231} 154-160 (1970).

\bibitem[26]{key-26}E. T. Whittaker and G. N. Watson, \emph{A Course
of Modern Analysis}, Cambridge (1996).

\bibitem[27]{key-27}M. Goppert-Mayer, \textquotedbl{}Uber Elementarakte
mit zwei Quantensprungen\textquotedbl{}, Ann. Physik \textbf{401}
273 (1931).

\bibitem[28]{key-28}P. Curie, \textquotedbl{}Lois experimentales
du magnetisme. Proprietes magnetiques des corps a diverses temperatures\textquotedbl{},
Ann. Chim. Phys. \textbf{5} 289 (1895). 

\bibitem[29]{key-29}P. Langevin, \textquotedbl{}Sur la theorie du
magnetisme\textquotedbl{}, J. Physique \textbf{4} 678 (1905).

\bibitem[30]{key-30}P. Langevin, \textquotedbl{}Magnetism et theorie
des electrons\textquotedbl{}, Ann. Chim. Phys. \textbf{5} 70 (1905).

\bibitem[31]{key-31}P. Debye, \textquotedbl{}Einige Resultate einer
kinetischen Theorie der Isolatoren\textquotedbl{}, Phys. Z. \textbf{13}
97 (1912).

\bibitem[32]{key-32}A. Micheli, G. Pupillo, H. P. Buchler and P.
Zoller, \textquotedbl{}Cold polar molecules in two-dimensional traps:
Tailoring interactions with external fields for novel quantum phases\textquotedbl{},
Phys. Rev. \textbf{A76} 043604 (2007)(1-25).

\bibitem[33]{key-33}L. Landau and E. Lifshitz, \emph{Course of Theoretical
Physics}, vol.8, \emph{Electrodynamics of Continuous Media}, Elsevier,
Oxford (1993).

\bibitem[34]{key-34}Ch. Kittel, \emph{Introduction to Solid State
Physics}, Wiley, NJ (2005). 

\bibitem[35]{key-35}J. C. Maxwell, \emph{Lehrbuch der Elektrizitat
und der Magnetismus}, vol. 1, Art. 328-330, Berlin (1983).

\bibitem[36]{key-36}K. W. Wagner, \textquotedbl{}Erklarung der dielektrischen
Nachwirkungsvorgange auf Grund Maxwellscher Vorstellungen\textquotedbl{},
Electr. Eng. (Archiv fur Elektrotechnik) \textbf{2} 371 (1914).

\bibitem[37]{key-37}K. W. Wagner, \emph{Die Isolierstoffe der Elektrotechnik},
H. Schering ed., Springer, Berlin (1924).

\bibitem[38]{key-38}R. W. Sillars, \textquotedbl{}The properties
of a dielectric containing semiconducting particles of various shapes\textquotedbl{},
J. Inst. Electr. Engrs. (London) \textbf{80} 378 (1937).

\bibitem[39]{key-39}A. von Hippel, \emph{Dielectrics and Waves},
Wiley, NY (1954).

\bibitem[40]{key-40}D. E. Aspnes, \textquotedbl{}Local-field effects
and effective-medium theory: a microscopic perspective\textquotedbl{},
Am. J. Phys. \textbf{50} 704 (1982).

\bibitem[41]{key-41}M. Apostol, S. Ilie, A. Petrut, M. Savu and S.
Toba, \textquotedbl{}Induced displacive transitions in heterogeneous
materials\textquotedbl{}, Eur. Phys. J. Appl. Phys. \textbf{59} 10401
(2012).

\bibitem[42]{key-42}P. W. Anderson, \textquotedbl{}Two comments on
the limits of validity of the P. R. Weisss theory of ferromagnetism\textquotedbl{},
Phys. Rev. \textbf{80} 922 (1950).

\end{thebibliography}
\end{document}